\documentclass[10pt]{iopart}

\usepackage{graphicx}
\usepackage{bm}
\usepackage{color}
\usepackage{threeparttable}

\begin{document}
\title{Hyperfine induced transitions probabilities from $4f^{14}5s5p~^3\mathrm{P}^o_{0,2}$ states in Sm-like ions}

\author{Fuyang Zhou$^{1}$,Jiguang Li$^1$,Yizhi Qu$^2$,Jianguo Wang$^1$}

\address{$^1$Data Center for High Density Physics, Institute of Applied Physics and Computational Mathematics,Beijing 100088, China\\
$^2$College of Material Sciences and Optoelectronic Technology, University of Chinese Academy of Sciences,Beijing 100049, China\\}

\ead{li\_jiguang@iapcm.ac.cn}

\begin{abstract}
The hyperfine induced $4f^{14}5s5p~^3\mathrm{P}^o_{0,2}~-~4f^{14}5s^2~^1\mathrm{S}_0$ transition probabilities for highly charged Sm-like ions are calculated in the framework of the multi-configuration Dirac-Hartree-Fock method. Electron correlation, the Breit interaction and quantum electrodynamical (QED) effects are taken into account.
For ions ranging from $Z=79$ to $Z=94$, $4f^{14}5s5p~^3\mathrm{P}^o_{0}$ is the first excited state, and the hyperfine induced transition is an dominant decay channel.
For the $4f^{14}5s5p~^3\mathrm{P}^o_{2}$ state, the hyperfine induced transition (HIT) rates of Sm-like ions with $Z=82-94$ are reported as well as the magnetic dipole (M1) $^3\mathrm{P}^o_2 ~-~ ^3\mathrm{P}^o_1$, the electric quadrupole (E2) $^3\mathrm{P}^o_2 ~-~ ^3\mathrm{P}^o_{0,1}$, and the magnetic quadrupole (M2) $^3\mathrm{P}^o_2 ~-~ ^1\mathrm{S}_0$ transition probabilities. It is found that M1 transition from the $4f^{14}5s5p~^3\mathrm{P}^o_2$ state is the most important decay channel in this range on $Z \ge 82$.
\end{abstract}



%
\maketitle
%
\ioptwocol

\section{Introduction}

The influence of the hyperfine interaction on lifetimes of metastable states was first noted by Bowen in a comment to a paper studying on the "forbidden" line of mercury by Huff and Houston~\cite{PhysRev.36.842}. 
The lifetimes of metastable states for $nsnp~^3\mathrm{P}_{0,2}$ states in the divalent atoms or ions were found to be significantly reduced by hyperfine induced transitions (HITs)~\cite{Johnson2011,Grumer2014}, especially for the $^3\mathrm{P}_0$ state because the single-photon $0-0$ transition is forbidden in absence of hyperfine interactions. 
In the presence of a finite nuclear spin, the hyperfine interaction introduces the mixings of wave functions between  $nsnp~^3P_{0,2}$ and  $^{1,3}P_{1}$ levels, opening single-photon electric dipole transitions to the ground state $ns^2~^1S_{0}$.
These HIT attracts much attention in view of its potential applications, for instance, developing atomic clocks~\cite{Takamoto2003,Wang2007,Rosenband2007}, diagnosing plasma parameters~\cite{0004-637X-500-1-507,Brage2002} and determining nuclear parameters~\cite{0295-5075-26-6-007,Toleikis2004}.
Since the pioneering work of Garstang~\cite{Garstang1962}, there are a number of theoretical studies of HITs of $nsnp~^3\mathrm{P}_{0,2}-ns^2~^1\mathrm{S}_0$ in the Be-like ($n=2$)~\cite{0004-637X-500-1-507,Marques1993,Cheng2008,Andersson2009,Jiguang}, the Mg-like ($n=3$)~\cite{0004-637X-500-1-507,Andersson2006,Kang2009,Kang2010}, and the Zn-like ($n=4$)~\cite{Liu2006,Andersson2008,Chen2011} isoelectronic sequences.
On the experimental side, Brage \emph{et al.} used a planetary nebula to determine the HIT rates of $2s2p~^3\mathrm{P}_{0}-2s^2~^1\mathrm{S}_0$ for Be-like $^{14}$N and $^{15}$N~\cite{Brage2002}.
 Schippers \emph{et al.} measured the lifetime of $2s2p~^3\mathrm{P}_{0}$ state for Be-like $^{47}$Ti and $^{33}$S~\cite{Schippers2007,Schippers2012}. Rosenband \emph{et al.} determined the lifetime of $3s3p~^3\mathrm{P}_{0}$ state for Mg-like $^{27}$Al~\cite{Rosenband2007}. 
The detail about HIT can be found in the review by Johnson~\cite{Johnson2011} and Grumer \emph{et al.}~\cite{Grumer2014}.

\begin{figure}[!ht]
\center
\includegraphics[scale=0.3]{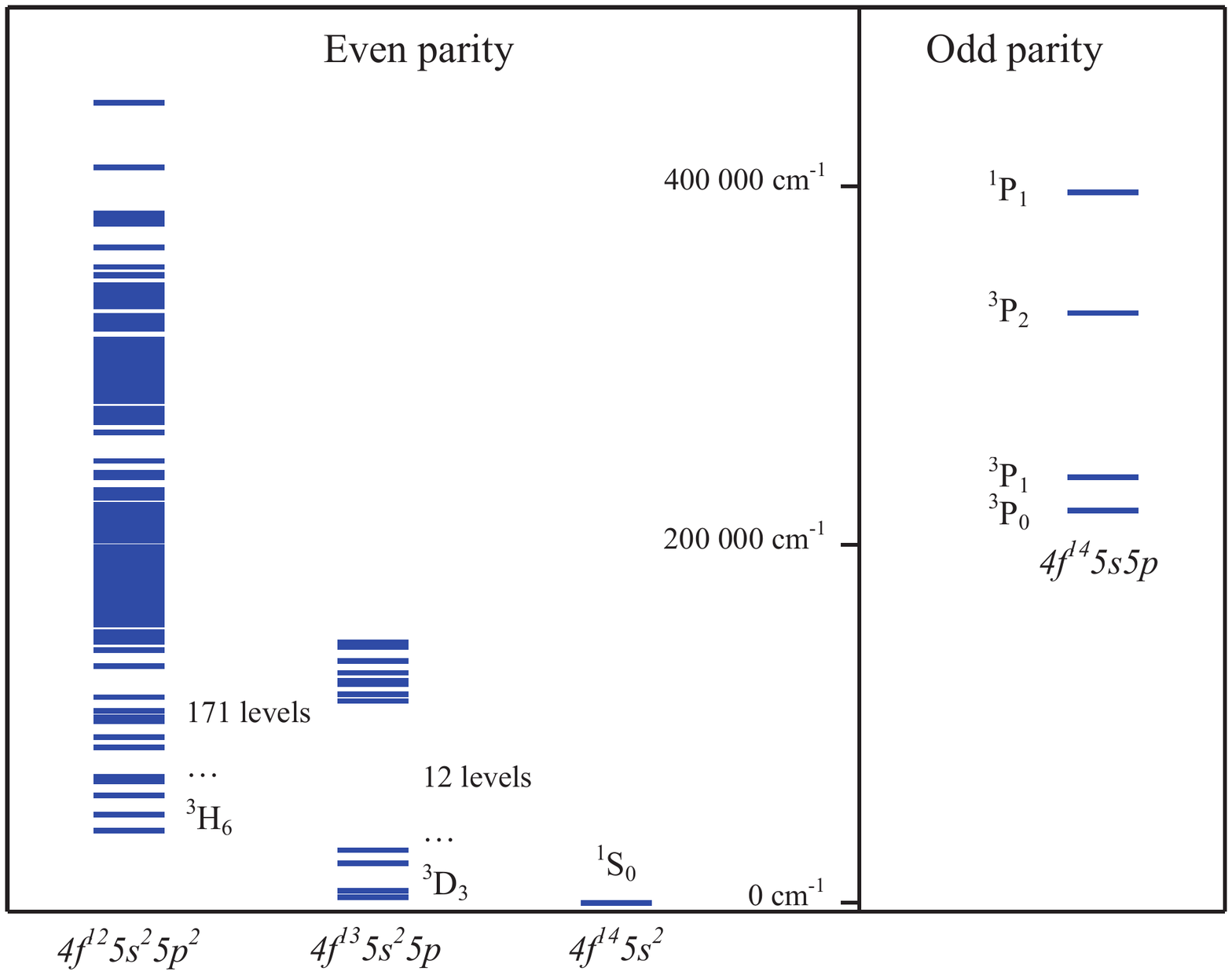}\\
\caption{\label{fig-W12} Scheme of the low-lying levels for W$^{12+}$. }
\end{figure}

\begin{figure}[!ht]
\center
\includegraphics[scale=0.3]{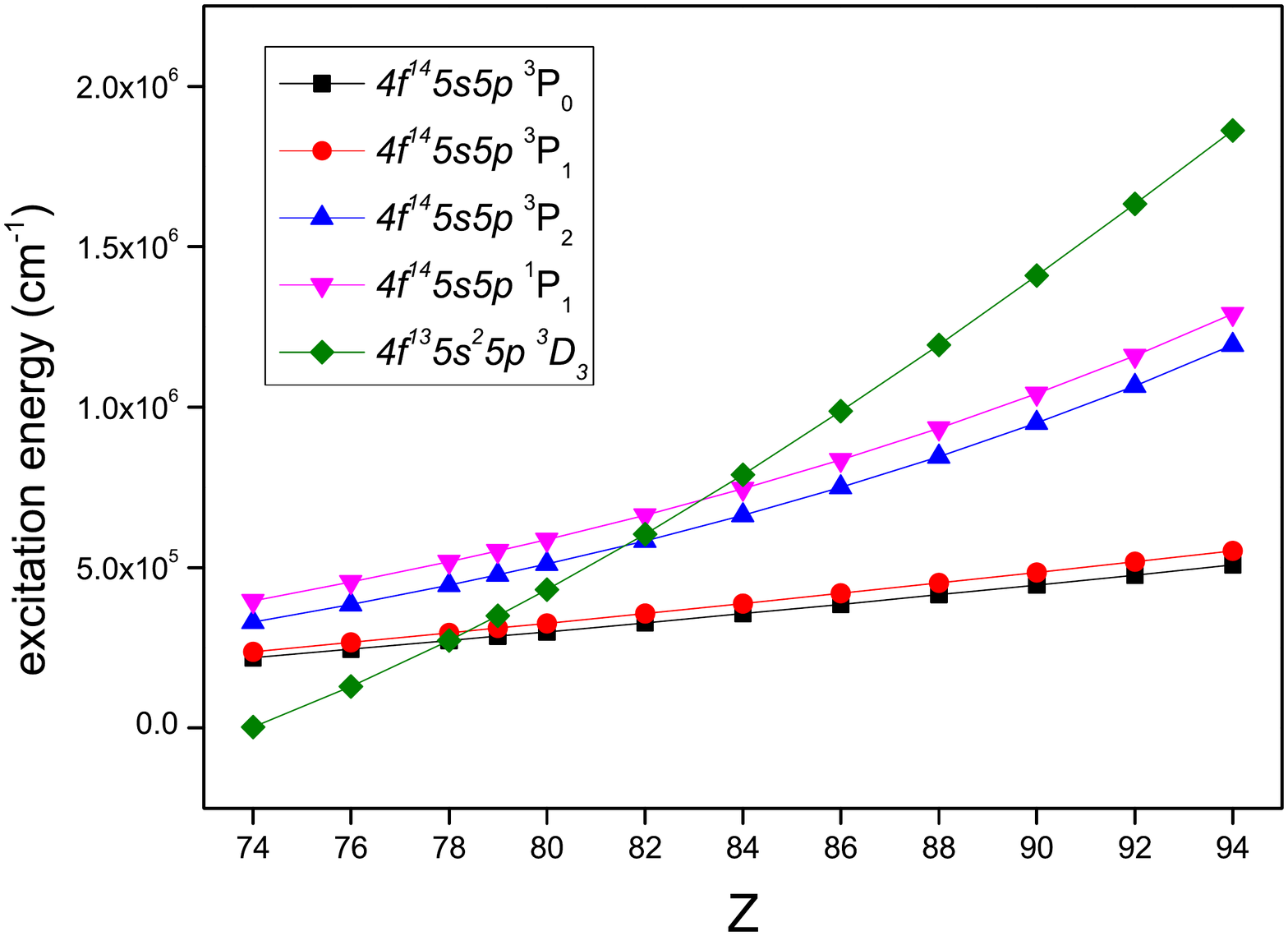}\\
\caption{\label{fig-S-conf} The excitation energies of $4f^{14}5s5p~^3\mathrm{P}_{0-2},~^1\mathrm{P}_1$ levels and the lowest one for configuration $4f^{13}5s^25p$ along the Sm-like isoelectronic sequence. }
\end{figure}

Here, we focus on the influence of HITs on lifetimes of the metastable  $4f^{14}5s5p~^3P_{0,2}$  states in Sm-like ions. 
The ground state of neutral Sm ($Z=62$) is [Kr]$4d^{10}5s^25p^64f^66s^2~^7\mathrm{F}_0$, and the ground states of Sm-like ions change several times with increasing atomic number due to the competition between $4f$, $5s$, $5p$ and $6s$ orbitals, eventually turn into the [Kr]$4d^{10}4f^{14}5s^2~^1\mathrm{S}_0$ state for W$^{12+}$ ($Z=74$) and heavier ions~\cite{Safronova2013}.
The scheme of the low-lying levels calculated in the Dirac-Hartree-Fock (DHF) approximation for W$^{12+}$ 
are shown in Fig.~\ref{fig-W12}. It is found that the spectra are much more complicated than other divalent atomic systems such as Mg- and Cd-like ion.
There are 12 states for configuration $4f^{13}5s^25p$ and 171 states for $4f^{12}5s^25p^2$, the energy levels of which overlap with those in odd parity configuration $4f^{14}5s5p$ due to the small energy difference between $4f$, $5s$, and $5p$ orbitals. 
The $4f^{14}5s5p~^3P_{0,2}$ states are not metastable  as the $5s5p~^3P_{0,2}$ states in other divalent atomic systems.
However, the energy difference between $4f$ and outmost $5s$ or $5p$ orbitals increases with $Z$, because the $4f$ shell is more tightly bound for ions with the higher degree of ionization~\cite{Safronova2013}.
In Fig.~\ref{fig-S-conf}, we present the calculated excitation energies of $4f^{14}5s5p~^3\mathrm{P}_{0,1,2}$, $^1\mathrm{P}_1$ states along the Sm-like isoelectronic sequence, as well as the excitation energy of $4f^{13}5s^25p$ $^3\mathrm{D}_3$ state which is the lowest level with respect to configurations with open $4f$ shell.
The curve describing the excitation energies of the $4f^{13}5s^25p$ $^3\mathrm{D}_3$ state crosses that for the $4f^{14}5s5p~^3P_0$ state around $Z = 79$, and also the one for the $4f^{14}5s5p~^3P_2$ state about $Z = 82$.
Therefore, the $4f^{14}5s5p~^3P_0$ and $^3\mathrm{P}_2$ levels become the metastable states when $Z \ge 79$ and $Z\ge 82$, respectively, and their lifetimes depend on probabilities of forbidden transitions.

In this work, we present an investigation on the hyperfine induced $4f^{14}5s5p~^3\mathrm{P}^o_{0,2}$ $-$  $4f^{14}5s^2~^1\mathrm{S}_0$ transitions for Sm-like ions between $Z=79$ to $Z=94$. Meanwhile, we calculate rates of the magnetic dipole (M1) $^3\mathrm{P}^o_2 ~-~ ^3\mathrm{P}^o_1$, the electric quadrupole (E2) $^3\mathrm{P}^o_2 ~-~ ^3\mathrm{P}^o_{0,1}$ and the magnetic quadrupole (M2) $^3\mathrm{P}^o_2 ~-~ ^1\mathrm{S}_0$ transitions, which are other important decay channels for the $4f^{14}5s5p~^3\mathrm{P}_2$ level of ions with $Z \ge 82$ besides the HIT. 

\section{Theory}

\subsection{The MCDHF method}
In the framework of the multi-configuration Dirac-Hartree-Fock (MCDHF) method, atomic state functions (ASFs) are represented as a linear combination of symmetry-adapted configuration state functions (CFSs)
\begin{eqnarray}
| {\Gamma J{M_J}} \rangle = \sum\limits_r {c_\gamma} | {{\gamma _r} J{M_J}} \rangle,
\end{eqnarray}
where $J$ and $M$ are the total angular momentum and the magnetic quantum numbers, respectively. $ \Gamma$ and $\gamma _r$ are the additional quantum number defining each ASF or CSF uniquely. Configuration mixing coefficients $c_\gamma$ are obtained through diagonalizing the Dirac-Coulumb Hamiltonian
\begin{eqnarray}
\label{eqn:2}
{H_{\rm{DC}}} = \sum\limits_{i = 1}^N {[c{\bm \alpha _i} \cdot {\bm p_i} + \left( {{\beta _i} - 1} \right){c^2} + V({r_i})]}  + \sum\limits_{i > j} {\frac{1}{{{r_{ij}}}}} .
\end{eqnarray}
Here, $V(r_i)$ is the monopole part of the electron-nucleus Coulomb interaction, and $\bm \alpha _i$ and $\beta_i$ are the Dirac matrices. In the relativistic self-consistent field  procedure, both the radial parts of Dirac orbitals and the mixing coefficients $c_\gamma$ are optimized for minimizing the energy of the atomic states concerned~\cite{Fischer1997}.

The Breit interaction in the low frequency approximation
and quantum electrodynamical (QED) effects including the vacuum polarization and the self-energy correction can be included in the subsequent relativistic configuration interaction (RCI) calculation~\cite{I.P.Grant2007,Jonsson2007,Jonsson2013}.

\subsection{Hyperfine induced transitions}

Hyperfine interactions couples  nuclear spin $I$ and electronic angular momentum $J$ to total angular momentum $F$, and only the $F$ and $M_F$ are good quantum numbers. The wave function of a hyperfine level can be written as
\begin{eqnarray}
| { F{M_F}} \rangle = \sum\limits_i {{h_{{\Gamma _i}{J_i}}}} |  {{\Gamma _i}{J_i}IF{M_F}} \rangle,
\end{eqnarray}
where $h_{{\Gamma _i}{J_i}}$ are the hyperfine mixing coefficients, and can be obtained in first-order perturbation theory  by
\begin{eqnarray}
h_{\Gamma _i J_i} = \frac{\langle \Gamma _i J_i IF M_F | H_{hfs} | \Gamma _0 J_0 I F M_F \rangle}{E_{\Gamma_0 J_0} - E_{\Gamma_i J_i}}.
\end{eqnarray}
Here, the subscript 0 labels the unperturbative level. The hyperfine interaction Hamiltonian $H_{hfs}$ is expressed as~\cite{Jonsson1996,Andersson20082}
\begin{eqnarray}
H_{hfs} = \sum_k {\bf T^{(k)}}  \cdot {\bf M^{(k)}},
\end{eqnarray}
where $\bf T^{(k)}$ and $\bf M^{(k)}$ are spherical tensor operators of rank $k$ acting on the electronic and the nuclear parts of the wave function, respectively. The hyperfine interaction matrix elements can be expressed in terms of reduced electronic and nuclear matrix elements as~\cite{Andersson2008}
\begin{eqnarray}
\label{eque7}
\begin{array}{l}
\langle {{\Gamma}{J}IF{M_F}} |{\bf T^{(k)}{\cdot}{\bf M^{(k)}}}| {{\Gamma _0}{J_0}IF{M_F}} \rangle\\
=(-1)^{I+J+F}\left\{
               \begin{array}{ccc}
                 I  & J & F \\
                 J_0& I & k \\
               \end{array}
             \right\}
\sqrt{2J+1}\sqrt{2I+1}\\
\times {\langle {{\Gamma}{J}} || {\bf T^{(k)}} || {{\gamma _0}{J_0}} \rangle}{\langle {{\Gamma}{I}} ||{\bf  {M^{(k)}}}|| {{\Gamma}{I}} \rangle}.
\end{array}
\end{eqnarray}
In this work, only the magnetic dipole ($k=1$) and the electric quadrupole ($k=2$) hyperfine interactions are included. The reduced matrix elements of the nuclear tensor operators are related to nuclear magnetic dipole moment $\mu_I$ and electric quadrupole moment $Q$ through
\begin{eqnarray}
\langle I ||{\bf M^{(1)}}|| I \rangle = \mu_I\sqrt{1+I^{-1}}
\end{eqnarray}
and
\begin{eqnarray}
\langle I ||{\bf M^{(2)}}|| I \rangle = \frac{Q}{2}\sqrt{\frac{(2I+3)(I+1)}{I(2I-1)}}.
\end{eqnarray}

The probability $A$ (in s$^{-1}$) for an electric dipole (E1) transition between two hyperfine levels can be written as~\cite{Andersson2009}
\begin{eqnarray}
A = \frac{{2.02613 \times {{10}^{18}}}}{{{\lambda ^3(2F'+1)}}} {\left| {\langle {\Gamma J IF} ||{D^{(1)}}|| {\Gamma 'J' I'F'} \rangle } \right|}^2.
\end{eqnarray}
Here, $\lambda$ is wavelength in \AA, and $D^{(1)}$ is the electric dipole tensor operator. The reduced matrix elements can be further simplified to
\begin{eqnarray}
\begin{array}{l}
 {\left\langle {\Gamma J IF} ||{D^{(1)}}|| {\Gamma 'J' I'F'} \right\rangle } = \sqrt{(2F+1)(2F'+1)}\\
 {\times (-1)^{J+I+F'+1}
                \left\{\begin{array}{ccc}
                       J  & F & I \\
                       F'& J' & 1 \\
                       \end{array}\right\}
\left\langle {\Gamma J} ||D^{(1)}|| {\Gamma 'J'} \right\rangle}.
\end{array}
\end{eqnarray}
Using Eq.(4), (10)and (11), the transition probability is given by
\begin{eqnarray}
\begin{array}{l}
A = \frac{{2.02613 \times {{10}^{18}}}}{{{\lambda ^3}}}(2F + 1)| {\sum\limits_{\Gamma J} {\sum\limits_{\Gamma 'J'} {{h_{\Gamma J}}{h_{\Gamma 'J'}}} } }\\
{{\left\{
               \begin{array}{ccc}
                 J  & F & I \\
                 F'& J' & 1 \\
               \end{array}
             \right\}\left\langle {\Gamma J} ||D^{(1)}|| {\Gamma 'J'} \right\rangle } |^2}.
\end{array}
\end{eqnarray}

In practical calculation, we only consider the hyperfine interactions between the states in the same configuration.
For example,  the hyperfine level belonging to the  $4f^{14}5s5p~^3\mathrm{P}_{0}$ state is given by
\begin{eqnarray}
\begin{array}{l}
| {^3\mathrm{P}^o_{0}IF{M_F}} \rangle^{(1)} =| {^3\mathrm{P}^o_{0}IF{M_F}} \rangle + h_{^{3}\mathrm{P}^o_{2}}| {^3\mathrm{P}^o_{2}IF{M_F}} \rangle \\
~+ \sum\limits_{S=0,1} {{h_{^{(2S+1)}\mathrm{P}^o_{1}}}} |  {^{(2S+1)}\mathrm{P}^o_{1}IF{M_F}} \rangle.
 \end{array}
\end{eqnarray}
The mixing between $4f^{14}5s5p~^3\mathrm{P}^o_{0,2}$ and $^{(2S+1)}\mathrm{P}^o_{1}$ ($S=0,1$) states gives rise to a single-photon E1 transition from the $4f^{14}5s5p~^3\mathrm{P}^o_{0,2}$ levels to the $4f^{14}5s^2~^1\mathrm{S}_0$ ground state.
The corresponding hyperfine induced $4f^{14}5s5p~^3\mathrm{P}^o_{0,2} $ $-$ $4f^{14}5s^2~^1\mathrm{S}_0$ transition rates can be given by
\begin{eqnarray}
\begin{array}{l}
A_{\rm{HIT}}(5s5p~^{3}\mathrm{P}^o_{J}-5s^2~^1\mathrm{S}_0) = \frac{{2.02613 \times {{10}^{18}}}}{{{3\lambda ^3}}} \\ \\
| {\sum\limits_{S=0,1} {{h_{^{(2S+1)}\mathrm{P}^o_{1}}}} }
{{\left\langle {5s^2~^1\mathrm{S}_0} ||D^{(1)}|| {5s5p~^{(2S+1)}\mathrm{P}^o_{1}} \right\rangle } |^2}.
\end{array}
\end{eqnarray}

Since the hyperfine mixing coefficients depend on nuclear parameters, it is difficult to investigate the trend of the HIT rates along the isoelectronic sequence. Therefore, we define a reduced HIT rate~\cite{0004-637X-500-1-507,Andersson2009,Kang2010}
\begin{eqnarray}
\label{eque13}
{A_{el}} = \frac{{A_{\rm{HIT}}^{\rm{M1}}}}{{\mu _I^2(1 + {I^{ - 1}})(2I + 1)(2J+1){{[W(I{J_0}JI;F1)]}^2}}}.\nonumber\\
\end{eqnarray}
Here, $W(IJ_0 JI; F1)$ is the $6j$-symbol as in Eq.(7) and $A_{\rm{HIT}}^{\rm{M1}}$ indicates that only the magnetic dipole hyperfine interaction was taken into account in $h_{^{(2S+1)}\mathrm{P}^o_{1}}$. It is clear that $A_{el}$ is independent on the nuclear parameters.

\subsection{Computational model}

The levels belonging to configuration $4f^{14}5s^2$ and $4f^{14}5s5p$ are optimized in two separate MCDHF calculations.
The configuration space is expanded by employing the active space approach.
We start from the Dirac-Hartree-Fock calculations, in which the occupied orbitals are optimized as spectroscopic.
In order to take into account the correlations between valence orbitals $4f$, $5s$ and $5p$, CSFs generated by single and double (SD) excitations from the valence orbitals to the virtual orbitals are included into the configuration space.
Due to stability problems in the relativistic self-consistent field procedure, the virtual orbitals are augmented layer by layer up to $n=8$, and only the outermost layer are optimized in each step. Each virtual orbital layer contains orbitals with the $s$, $p$, $d$, $f$ and $g$ symmetries.
In Table~\ref{table1} the virtual orbitals are listed by angular symmetry and the number of orbitals for each symmetry, being enclosed in quotation marks to avoid confusion with spectroscopic orbitals. For example, ``$2spdfg$'' stands for two $s$, $p$, $d$, $f$ and $g$ virtual orbitals. The core-valence (CV) and core-core (CC) correlations involving $4s$, $4p$ and $4d$ shells are taken into account in the following RCI calculations. In this step CFSs generated by SD excitations from these three core orbitals to all  virtual orbitals are included. 
By this step the calculations had converged, but in order to include spin-polarization  in deep $s$-subshell~\cite{J1993Large,PhysRevA.77.042509} we add the configurations  generated by  single excitations from 1$s$, 2$s$ and 3$s$, to virtual $s$-orbitals in the last step.
Furthermore, the Breit interaction and QED effects are considered in the RCI computations.
 
\section{Results and discussion}

Taking Au$^{17+}$ ($Z=79$) as an example, we present in Table~\ref{table1} the excitation energies of $4f^{14}5s5p~^{1,3}\mathrm{P}^o_{J}$ states as functions of the computational models, as well as the available experimental and theoretical values.
In Curtis' calculation~\cite{Curtis1986}, only the configurations $4f^{14}5s^2$, $4f^{14}5s5p$, $4f^{13}5s^25p$, and $4f^{13}5s5p^2$ are included, and the excitation energy of $4f^{14}5s5p~^{3}\mathrm{P}^o_{1}$ is about 8000 cm$^{-1}$ lower than the experimental value~\cite{Kaufman1990}.
The present multiconfiguration calculation give much better results.
It was found that the main valence correlations were captured by three virtual orbital layers, which make larger contribution to the excitation energies than the CV and CC correlations. The effects of the Breit interaction and QED on the excitation energies are similar to the CV and CC correlations.
\begin{table}[!ht]
\centering
\caption{\label{table1}%
The calculated excitation energies (in cm$^{-1}$) of $4f^{14}5s5p~^{1,3}\mathrm{P}^o_{J}$ states for the Sm-like Au ion. DHF represents the uncorrelated Dirac-Hartree-Fock calculation. ``$nspdfg$'' stands for the virtual orbital set. CV \& CC indicate RCI computations taking core-valence and core-core correlations into account.
}
\begin{threeparttable}
\begin{tabular}{ccccc}
\hline
\hline
Model                     & $^3\mathrm{P}^o_0$ & $^3\mathrm{P}^o_1$ & $^3\mathrm{P}^o_2$ & $^1\mathrm{P}^o_1$ \\
\hline
DHF                        & 285869    & 311110    & 476907     & 551998 \\
``$1spdfg$''               & 296863    & 319045    & 488005     & 549689 \\
``$2spdfg$''               & 299017    & 320587    & 490166     & 549180 \\
``$3spdfg$''               & 299292    & 320731    & 490367     & 548763 \\
+ CV \& CC                & 297896    & 319459    & 489498     & 548367 \\
+Breit                         & 299450    & 320918   &  489125     &  548024 \\
+QED                        & 296660    & 318132    & 486483     & 545359 \\
Curtis~\cite{Curtis1986}   & 285669    & 310705    & 474659     & 549501 \\
Expt.~\cite{Kaufman1990}   &           & 318878\tnote{a} &&\\
\hline
\hline
\end{tabular}
 \begin{tablenotes}
        \footnotesize
        \item[a] it is converted from the measured wavelength 31.36 nm.
 \end{tablenotes}
\end{threeparttable}
\end{table}

\begin{table}[!ht]
\centering
\caption{\label{table1+}%
The line strengths of the $^{1,3}\mathrm{P}^o_1~-~^1\mathrm{S}_0$ E1 transitions for the Sm-like Au ion. $S_B$ and $S_C$ stand for the line strengths in Babushkin and Coulomb gauges, respectively.}
\begin{tabular}{cccccc}
\hline
\hline
                         & \multicolumn{2}{c}{$^{1}\mathrm{P}^o_1~-~^1\mathrm{S}_0$} && \multicolumn{2}{c}{$^{3}\mathrm{P}^o_1~-~ ^1\mathrm{S}_0$}\\
\cline{2-3}\cline{5-6} Model   & $S_C$     &$S_B$      &&$S_C$       &$S_B$\\
 \hline
DHF                    & 1.028    & 1.274     && 0.165     & 0.237 \\
``$1spdfg$''           & 0.959    & 1.000     && 0.194     & 0.202 \\
``$2spdfg$''           & 0.909    & 0.944     && 0.189     & 0.198 \\
``$3spdfg$''           & 0.895    & 0.930     && 0.186     & 0.196 \\
+ CV \& CC             & 0.902    & 0.933     && 0.179     & 0.199 \\
+Breit                      &  0.907    &  0.937    &&  0.180     &  0.198          \\
+QED          & 0.919    & 0.937     && 0.184     & 0.198 \\
 \hline
 \hline
\end{tabular}
\end{table}

Table~\ref{table1+} presents the line strengths of the $^{1,3}\mathrm{P}^o_1~-~^1\mathrm{S}_0$ E1 transitions in the Babushkin and Coulomb gauges for Au$^{17+}$, respectively. Convergence trends of the line strengths are similar to the excitation energies. In addition, with the expansion of the configuration space, the agreement between the two gauges is significantly improved. After including the Breit interaction and QED effects, the difference in line strengths between these two gauge is only about 2\% for the $^1\mathrm{P}^o_1$ - $^1\mathrm{S}_0$ transition and 8\% for the $^3\mathrm{P}^o_1$ - $^1\mathrm{S}_0$ transition. The line strength in the Babushkin gauge are used to calculate the HIT rates, since it is relatively insensitive to electron correlation effects~\cite{PhysRevA.92.052505}.

\subsection{Hyperfine induced $4f^{14}5s5p~^3\mathrm{P}^o_0 ~-~ 4f^{14}5s^2~^1\mathrm{S}_0$ transition probability}

\begin{table*}
\centering
\caption{\label{table2}
Transition energies ($\triangle E$) in cm$^{-1}$, probabilities ($A_{\rm{HIT}}$) and reduced rates ($A_{el}$) in $s^{-1}$ of hyperfine induced transition $4f^{14}5s5p$ $^3\mathrm{P}^o_0$ $-$ $4f^{14}5s^2$ $^1\mathrm{S}_0$ for Sm-like ions with $Z=79-94$. The uncertainty in the nuclear parameters is given in parentheses.
}
\begin{tabular}{cclcrr}
 \hline
 \hline
 Nucleus  & I   & $\mu_I$      & $\triangle E$ & $A_{el}$  &$A_{\rm{HIT}}$\\
 \hline
$^{185}$Au & 5/2 & 2.17(2)     & 296660        & 207.976   & 1371.074       \\
$^{197}$Au & 3/2 & 0.145746(9) & 296660        & 207.976   & 7.391         \\
$^{199}$Hg & 1/2 & 0.5058855(9)& 310469        & 255.922   & 196.287       \\
$^{203}$Hg & 5/2 & 0.84895(13) & 310469        & 255.922   & 258.226       \\
$^{205}$Pb & 5/2 & 0.7117(4)   & 338513        & 382.247   & 271.061       \\
$^{209}$Pb & 9/2 & 1.4735(16)  & 338513        & 382.247   & 1014.366       \\
$^{207}$Po & 5/2 & 0.79(6)     & 367140        & 562.262   & 491.271       \\
$^{209}$Po & 1/2 & 0.68(8)     & 367140        & 562.262   & 779.970       \\
$^{209}$Rn & 5/2 & 0.8388(4)   & 394824        & 812.550   & 800.378       \\
$^{211}$Rn & 1/2 & 0.601(7)    & 394824        & 812.550   & 880.482       \\
$^{211}$Ra & 5/2 & 0.878(4)    & 424640        & 1162.946  & 1255.096      \\
$^{223}$Ra & 3/2 & 0.271(2)    & 424640        & 1162.946  & 141.861       \\
$^{227}$Ra & 3/2 & 0.404(2)    & 424640        & 1162.946  & 315.248       \\
$^{229}$Th & 5/2 & 0.46(4)     & 455129        & 1642.883  & 486.448       \\
$^{233}$U  & 5/2 & 0.59(5)     & 486888        & 2305.913  & 1123.768      \\
$^{236}$U  & 7/2 & 0.38(3)     & 486888        & 2305.913  & 429.823       \\
$^{239}$Pu & 1/2 & 0.203(4)    & 518392        & 3208.076  & 398.292       \\
$^{241}$Pu & 5/2 & 0.683(15)   & 518392        & 3208.076  & 2095.146      \\
 \hline
 \hline
\end{tabular}
\end{table*}

\begin{figure}[!ht]
\centering
\includegraphics[scale=0.3]{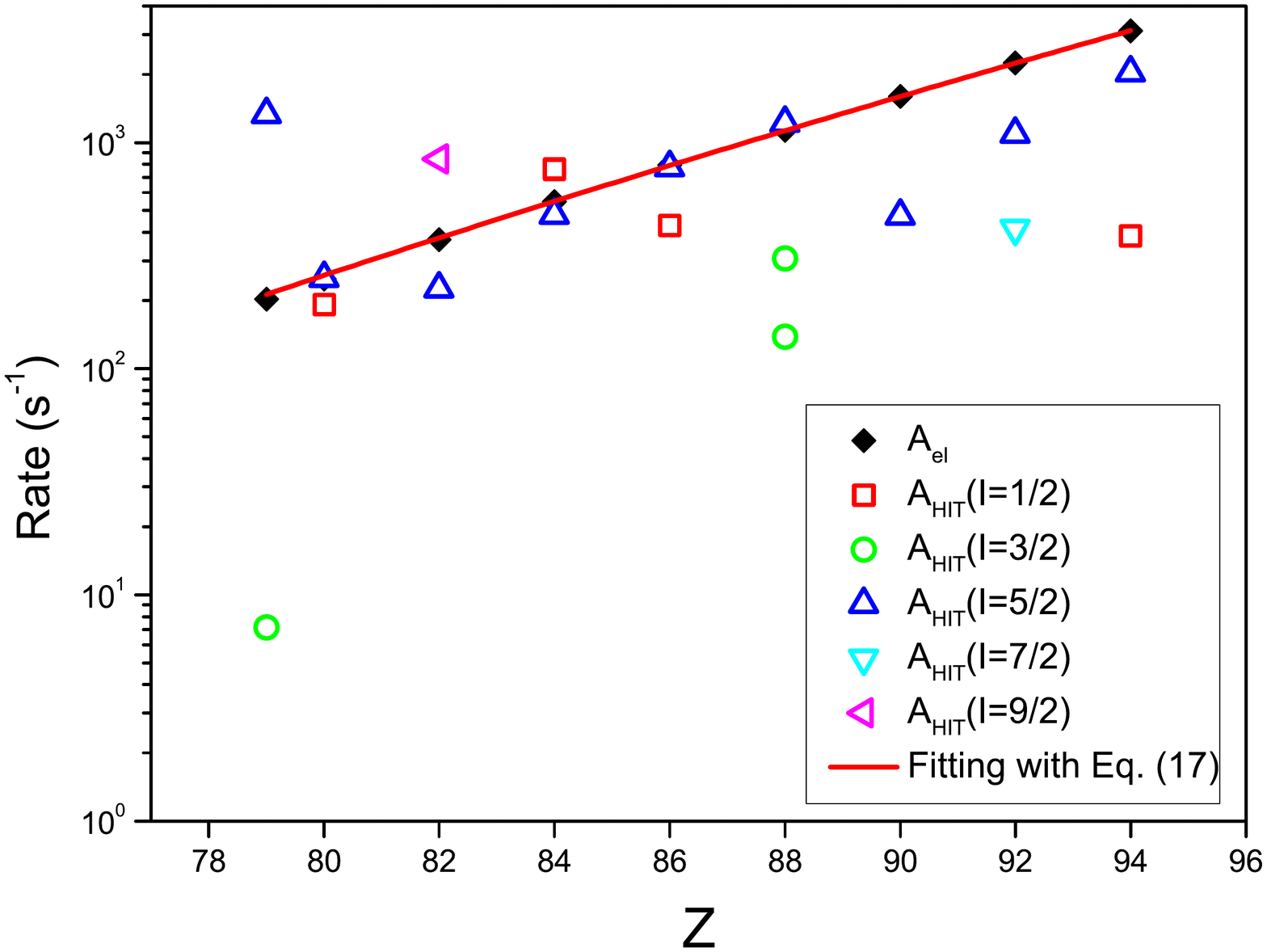}\\
\caption{\label{fig-3P0} The rates in $s^{-1}$ of hyperfine induced $^3\mathrm{P}^o_0$ - $^1\mathrm{S}_0$ electric dipole transition $A_{\rm{HIT}}$  and reduced hyperfine induced transition $A_{el}$, as well as the fitting curve by using Eq. (14)}
\end{figure}

In Table~\ref{table2} the hyperfine induced $4f^{14}5s5p$ $^3\mathrm{P}^o_0 ~-~ 4f^{14}5s^2 ~^1\mathrm{S}_0$ transition probabilities ($A_{\rm{HIT}}$) and corresponding transition energies ($\Delta E$) for Sm-like ions with $Z=79-94$ are displayed. The nuclear parameters are taken from Ref~\cite{Stone2005}.
For these ions the $^3\mathrm{P}^o_0$ level is the first excited state. Therefore, the hyperfine induced transition is the unique single-photon decay channel, and the lifetime of this state is determined by the HIT rates. As can be seen from this table, the HIT rates depend on the nuclear properties to a large extent. For example, the HIT probabilities for $^{223}$Ra (I=3/2, $\mu_I$=0.271) is 137.937 $s^{-1}$, compared with 306.553 $s^{-1}$ for the isotope $^{227}$Ra (I=3/2, $\mu_I$=0.404).
In order to investigate the trend of the HIT probabilities along the isoelectronic sequence, the `reduced' hyperfine-induced transition rates ($A_{el}$) defined in Eq.~(12) are illustrated in Fig.~\ref{fig-3P0}.
For $4f^{14}5s5p~^3\mathrm{P}_0$ state, there is no electric quadrupole hyperfine interaction between $^3\mathrm{P}_0$ and $^{1,3}\mathrm{P}_1$ states. Hence, the HIT rate can be expressed as
\begin{eqnarray}
A_{\rm{HIT}}(^3\mathrm{P}_0 - ^1\mathrm{S}_0)=\mu ^2_I(1+I^{-1})A_{el}(^3\mathrm{P}_0 - ^1\mathrm{S}_0),
\end{eqnarray}
which factorizes the HIT rate into the nuclear and the electronic parts.
The reduced HIT rates $A_{el}$  are independent of the nuclear parameters and have a smooth behaviour along the isoelectronic sequence.
Furthermore, the reduced HIT rates can be fitted as a power function of $Z$ as
\begin{eqnarray}
A_{el}(^3\mathrm{P}_0 - ^1\mathrm{S}_0)=1.0198\times 10^{-27}Z^{15.4569}.
\end{eqnarray}
The fitting curve is also depicted in Fig.~\ref{fig-3P0}, which is in good agreement with the \textit{ab initio} calculation.

\subsection{Decay of the $4f^{14}5s5p~^3\mathrm{P}^o_2$ state}

\begin{table*}[!ht]
\centering
\caption{\label{table3}%
Transition energies ($\triangle E$) in cm$^{-1}$, probabilities ($A_{\rm{HIT}}$) and reduced rates ($A_{el}$) in $s^{-1}$ of hyperfine induced transition $4f^{14}5s5p$ $^3\mathrm{P}^o_2$ - $4f^{14}5s^2$ $^1\mathrm{S}_0$ for Sm-like ions with $Z=82-94$. Numbers in square brackets represent powers of 10.
}
\begin{tabular}{ccllcccc}
 \hline
 \hline
 Nucleus   & I   & $\mu_I$      & $Q$    & $\triangle E$ & $A_{el}$ & F   &$A_{\rm{HIT}}$  \\
 \hline
$^{205}$Pb & 5/2 & 0.7117(4)    &0.23(4) &   592380   & 6.188[3] & 1/2 & 0       \\
           &     &              &        &            &          & 3/2 & 8.209[2]\\
           &     &              &        &            &          & 5/2 & 1.628[3]\\
           &     &              &        &            &          & 7/2 & 1.660[3] \\
           &     &              &        &            &          & 9/2 & 0        \\
$^{209}$Po & 1/2 & 0.68(8)      &        &  671560    & 1.065[4] & 3/2 & 7.389[3] \\
           &     &              &        &            &          & 5/2 & 0        \\
$^{209}$Rn & 5/2 & 0.8388(4)    & 0.31(3)&  756856    & 1.804[4] & 1/2 & 0        \\
           &     &              &        &            &          & 3/2 & 3.335[3] \\
           &     &              &        &            &          & 5/2 & 6.601[3] \\
           &     &              &        &            &          & 7/2 & 6.710[3] \\
           &     &              &        &            &          & 9/2 & 0        \\
$^{211}$Ra & 5/2 & 0.878(4)     &0.46(5) & 852112     & 3.045[4] & 1/2 & 0        \\
           &     &              &        &            &          & 3/2 & 6.263[3] \\
           &     &              &        &            &          & 5/2 & 1.228[4] \\
           &     &              &        &            &          & 7/2 & 1.231[4] \\
           &     &              &        &            &          & 9/2 & 0        \\
$^{229}$Th & 5/2 & 0.46(4)      &4.3(9)  & 956641     &	5.105[4] & 1/2 & 0        \\
           &     &              &        &            &          & 3/2 & 6.116[3] \\
           &     &              &        &            & 	     & 5/2 & 7.794[3]  \\
           &     &              &        &            & 	     & 7/2 & 3.286[3]  \\
           &     &              &        &            & 	     & 9/2 & 0           \\
$^{233}$U  & 5/2 & 0.59(5)      &3.663(8)& 1072110    &	8.527[4] & 1/2 & 0           \\
           &     &              &        &            & 	     & 3/2 & 1.302[4]    \\
           &     &              &        &            &          & 5/2 & 1.906[4]  \\
           &     &              &        &            &          & 7/2 & 1.132[4]  \\
           &     &              &        &            &          & 9/2 & 0       \\
$^{241}$Pu & 5/2 & 0.683(15)    &6(2)    & 1197904    & 1.416[5] & 1/2 & 0       \\
           &     &              &        &            &          & 3/2 & 3.426[4] \\
           &     &              &        &            &          & 5/2 & 4.575[4]  \\
           &     &              &        &            &          & 7/2 & 2.186[4]  \\
           &     &              &        &            &          & 9/2 & 0        \\
 \hline
 \hline
\end{tabular}
\end{table*}

\begin{figure}[!ht]
\center
\includegraphics[scale=0.3]{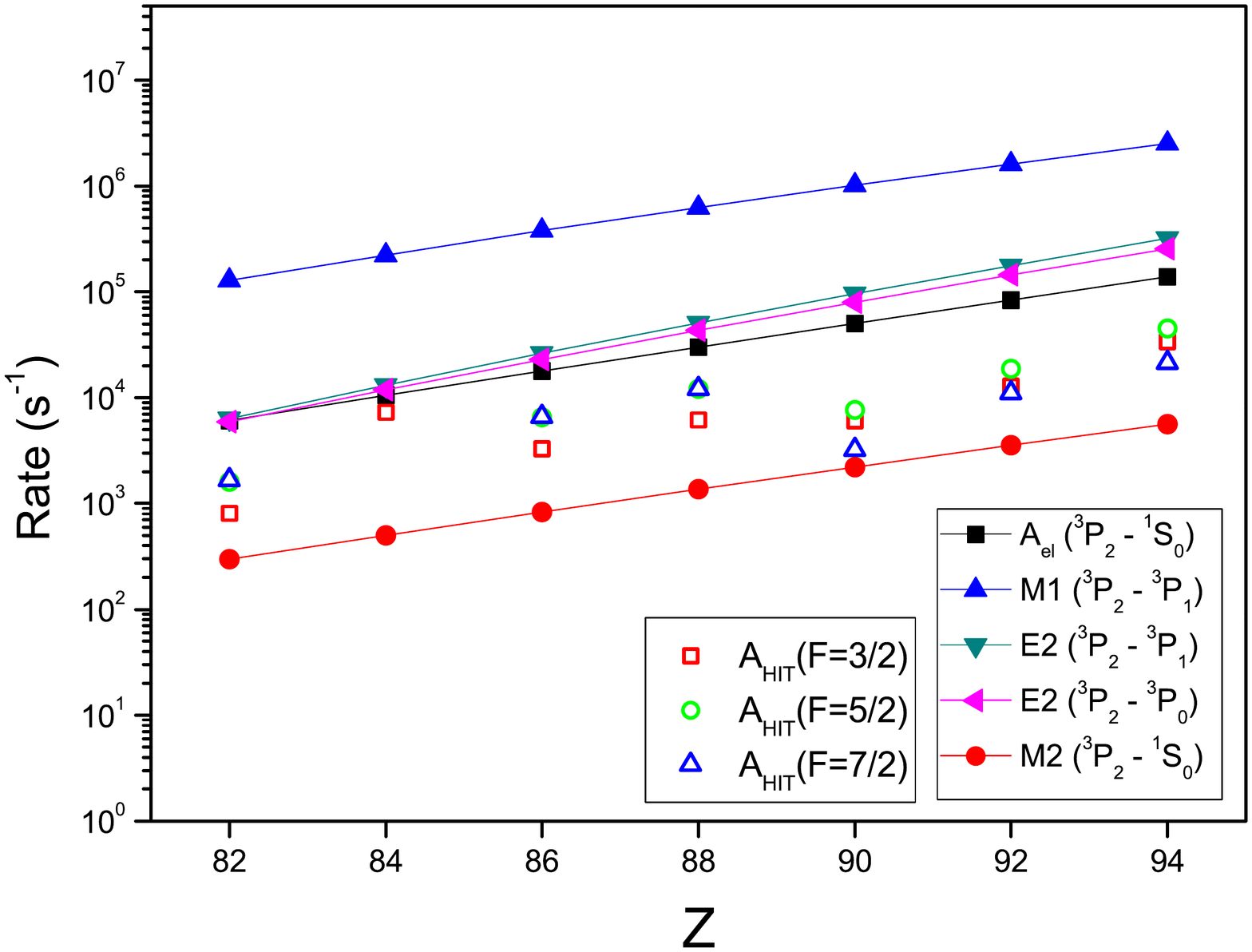}\\
\caption{\label{fig-3P2} The rates in $s^{-1}$ of hyperfine induced $^3\mathrm{P}^o_2$ - $^1\mathrm{S}_0$ electric dipole transition $A_{\rm{HIT}}$ (scattering symbols), reduced hyperfine induced transition ($A_{el}$), magnetic dipole transition (M1), electric quadrupole transitions (E2) and magnetic quadrupole transition (M2).}
\end{figure}

The HIT probabilities ($A_{\rm{HIT}}$) of the $^3\mathrm{P}^o_2$ level and corresponding transition energies ($\triangle E$) for Sm-like ions with $Z=82-94$ are presented in Table~\ref{table3}.
Apart from the hyperfine-induced transition, the $^3\mathrm{P}^o_2$ state can decay through the magnetic dipole (M1) transition ($^3\mathrm{P}^o_2$ $-$ $^3\mathrm{P}^o_1$), the electric quadrupole (E2) transitions ($^3\mathrm{P}^o_2~-~^3\mathrm{P}^o_{0,1}$) and the magnetic quadrupole (M2) transition ($^3\mathrm{P}^o_2$ $-$ $^1\mathrm{S}_0$). To show the competition among these decay channels, the trend of the reduced HIT probabilities as well as the M1, M2 and E2 transition probabilities of this state along the isoelectronic sequence are illustratively presented in Fig.~\ref{fig-3P2}. The HIT probabilities for the isotopes of each element concerned are plotted with scattering symbols. It is found that the M1 transition is dominant for these ions, and the HIT probabilities have the same order of magnitude as those two E2 transitions ($^3\mathrm{P}^o_2~-~^3\mathrm{P}^o_{0,1}$), but also depend on the nuclear parameters.

\begin{figure}
\center
\includegraphics[scale=0.3]{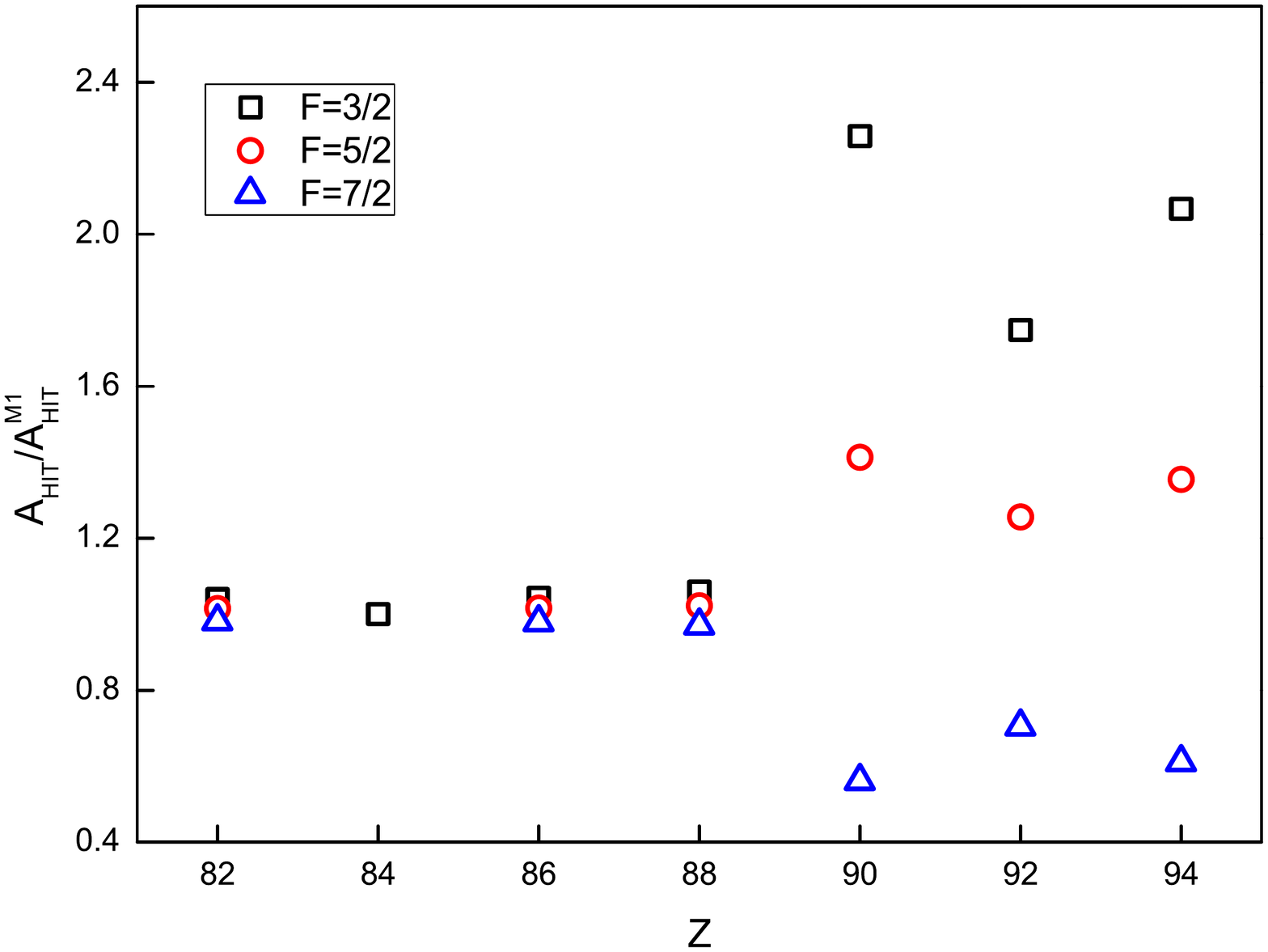}\\
\caption{\label{fig-ratio}   The ratio between $A_{\rm{HIT}}$ and $A^{\rm{M1}}_{\rm{HIT}}$. Only the magnetic dipole hyperfine interaction were considered in $A^{\rm{M1}}_{\rm{HIT}}$}
\end{figure}

As mentioned above, the electric quadrupole hyperfine interaction is not taken into account in $A_{el}$.
In order to show the contribution from the electric quadrupole hyperfine interaction to the HIT rate for the $^3\mathrm{P}^o_2$ state, the ratios between the hyperfine induced transition probabilities $A_{\rm{HIT}}$ and the rates $A^{\rm{M1}}_{\rm{HIT}}$, where the latter only includes the magnetic dipole hyperfine interaction, are presented in Fig.~\ref{fig-ratio}. 
It is clear that the electric quadrupole hyperfine interaction is important to the HIT rate only for isotopes with larger nuclear electric quadrupole moment $Q$ compared with magnetic dipole moment $\mu_I$. For example, for $^{229}$Th with $Q=4.3(9)$ and $\mu_I =0.46(4)$ the electric quadrupole hyperfine interaction changes the HIT rate ($F=3/2$) by a factor of about 2.
For isotope $^{209}$Po with nuclear spin $I=1/2$, however, the HIT only depends on the magnetic dipole hyperfine interaction.

\section{Summary}

With increasing atomic number $Z$, the ground state of Sm-like ions become [Kr]$4d^{10}4f^{14}5s^2~^1\mathrm{S}_0$, and the $4f^{14}5s5p~^3\mathrm{P}^o_0$ and $^3\mathrm{P}^o_2$ states turn into metastable states for the ions with $Z\geq79$ and $Z\geq82$, respectively. Using the multi-configuration Dirac-Hartree-Fock method, we calculate the hyperfine induced $4f^{14}5s5p~^3\mathrm{P}^o_{0,2}-4f^{14}5s^2~^1\mathrm{S}_0$ transition rates for these ions. For $4f^{14}5s5p~^3\mathrm{P}^o_2$ state, the probabilities of other important decay channels including the M1, M2 and E2 transitions are also reported. For the first excited state $4f^{14}5s5p~^3\mathrm{P}^o_0$, the hyperfine induced transition is a dominant single-photon decay channel, and thus significantly impact the lifetime of this state. A fitting formula in $Z$ for the electronic part of HIT rate is further derived in order to estimate the hyperfine induced transition rate for any isotope along the isoelectronic sequence. For the other metastable state, it is shown that the M1 transition is the most important decay channel for these ions with $Z \ge 82$.

\section{Acknowledgments}
This work is supported by the National Natural Science Foundation of China (Grant No. 11404025), NSAF (Grants No. U1330117) and China Postdoctoral Science Foundation (Grant No.2014M560061).

\section*{References}
\bibliography{smlike-final}
\bibliographystyle{unsrt}



\end{document}